\documentclass[conference]{IEEEtran}
\IEEEoverridecommandlockouts
\usepackage{cite}
\usepackage{amsmath,amssymb,amsfonts}
\usepackage{algorithmic}
\usepackage{graphicx}
\usepackage{textcomp}
\usepackage{xcolor}
\usepackage{multirow}
\usepackage{boldline}
\usepackage{amsmath}
\usepackage{amssymb}
\usepackage{bm}
\usepackage{hyperref}
\newcommand{\A}{\mathcal{A}}
\newcommand{\bmy}{\bm{y}}
\newcommand{\bms}{\bm{s}}
\newcommand{\bmh}{\bm{h}}
\newcommand{\Brace}[1]{\left\{#1\right\}}

\newcommand{\SLVD}{\text{SLVD}}

\newcommand{\SSV}{\text{SSV}}
\newcommand{\trace}{\text{trace}}
\newcommand{\List}{\text{list}}
\newcommand{\E}{\mathsf{E}}

\DeclareMathOperator*{\IEE}{IEE}

\usepackage{array}
\newcolumntype{P}[1]{>{\centering\arraybackslash}p{#1}}
    
\begin{document}

\title{High-Rate Convolutional Codes with CRC-Aided List Decoding for Short Blocklengths\\
\thanks{This research is supported by National Science Foundation (NSF) grant CCF-2008918. Any opinions, findings, and conclusions or recommendations expressed in this material are those of the author(s) and do not necessarily reflect views of NSF.}
}

% \author{\IEEEauthorblockN{Wenhui Sui}
% \and
% \and
% \and
% \and
% \and
\author{\IEEEauthorblockN{
Wenhui Sui,
Hengjie Yang, 
Brendan Towell,
Ava Asmani, 
and
Richard D. Wesel}
\IEEEauthorblockA{Department of Electrical and Computer Engineering\\
University of California, Los Angeles}
Email: \{wenhui.sui, hengjie.yang, brendan.towell, ava24, wesel\}@ucla.edu
}

\maketitle 

\begin{abstract}
Recently, rate-$1/\omega$ zero-terminated and tail-biting convolutional codes (ZTCCs and TBCCs) with cyclic-redundancy-check (CRC)-aided list decoding have been shown to closely approach the random-coding union (RCU) bound for short blocklengths. This paper designs CRC polynomials for rate-$(\omega-1)/\omega$ CCs with short blocklengths, considering both the ZT and TB cases. The CRC design seeks to optimize the frame error rate (FER) performance of the code resulting from the concatenation of the CRC code and the CC. Utilization of the dual trellis proposed by Yamada \emph{et al.} lowers the complexity of CRC-aided serial list Viterbi decoding (SLVD) of ZTCCs and TBCCs. CRC-aided SLVD of the TBCCs closely approaches the RCU bound at blocklength of $128$.

\end{abstract}

% \begin{IEEEkeywords}

% \end{IEEEkeywords}

\section{Introduction}

The structure of concatenating a convolutional code (CC) with a cyclic redundancy check (CRC) code has been a popular paradigm since 1994 when it was proposed in the context of hybrid automatic repeat request (ARQ) \cite{Rice1994}. It was subsequently adopted in the cellular communication standards of both 3G \cite{3GPP2006} and 4G LTE \cite{3GPP2018}. In general, the CRC code serves as an outer error-detecting code that verifies if a codeword has been correctly received, whereas the CC serves as an inner error-correcting code to combat channel errors.

Recently, there has been a renewed interest in designing powerful short blocklength codes. This renewed interest is mainly driven by the development of finite blocklength information theory by Polyanskiy \emph{et al.}, \cite{Polyanskiy2010} and the stringent requirement of ultra-reliable low-latency communication (URLLC) for mission-critical IoT (Internet of Things) services \cite{Ji2018}. In \cite{Polyanskiy2010}, Polyanskiy \emph{et al.} developed a new achievability bound known as the random-coding union (RCU) bound and a new converse bound, known as the meta-converse (MC) bound. Together, these two bounds characterize the error probability of the best short blocklength code of length $n$ with $M$ codewords. The URLLC for mission-critical IoT requires that the time-to-transmit latency is within 500 $\mu s$ while maintaining a frame error rate (FER) less than $10^{-5}$.

Several short blocklength code designs have been proposed in literature. Important examples include the tail-biting (TB) convolutional codes decoded with the wrap-around Viterbi algorithm (WAVA) \cite{Gaudio2017}, extended Bose-Chaudhuri-Hocquenghem (BCH) codes under ordered statistics decoding \cite{Coskun2019, Yue2021}, non-binary low-density parity-check (LDPC) codes \cite{Dolecek2014}, non-binary turbo codes \cite{Liva2013}, and polar codes under CRC-aided successive-cancellation list decoding \cite{Tal2015}. Recent advances also include the polarization-adjusted convolutional codes by Ar\i kan\cite{Arikan2019}. As a comprehensive overview, Co{\c s}kun \emph{et al.} \cite{Coskun2019} surveyed most of the contemporary short blocklength code designs in the recent decade. We refer the reader to \cite{Coskun2019} for additional information.  %Srinivasan \emph{et al.} discussed applying the symbol-by-symbol decoding algorithm derived from \cite{Hartmann1976} on a high-rate dual trellis, which has a Fourier transform relation with the \emph{primal} trellis.
% a (Fourier-transform-based) dualization of trellises in the context of Hartmann-Rudolph decoding

In \cite{Yang2022}, Yang \emph{et al.} proposed the CRC-aided CC as a powerful short blocklength code for binary-input (BI) additive white Gaussian noise (AWGN) channels. In \cite{Yang2022}, the convolutional encoder of interest is either zero-terminated (ZT) or TB with rate-$1/\omega$. A good CRC-aided CC is constructed by searching for the \emph{distance-spectrum optimal} (DSO) CRC polynomial for a given convolutional encoder and then concatenating the DSO CRC polynomial with the convolutional encoder. 

The nature of the concatenation naturally permits the use of serial list Viterbi decoding (SLVD), an efficient algorithm originally proposed by Seshadri and Sundberg \cite{Seshadri1994}. Yang \emph{et al.} showed that the expected list rank $\E[L]$ of the SLVD of the CRC-aided CC is small at a target low error probability, thus achieving a low average decoding complexity. Yang \emph{et al.} demonstrated that several concatenated codes generated by the DSO CRC polynomial and the TBCC, or in short, CRC-TBCCs, approach the RCU bound. In \cite{Schiavone2021}, Schiavone extended this line of work by looking at the parallel list Viterbi decoding with a bounded list size. However, these works did not consider rate-$(\omega-1)/\omega$ CCs. It remains open whether this framework can be extended to CCs of an arbitrary rate such that the resulting concatenated code can approach the RCU bound at a low decoding complexity.

In this paper, we consider designing good CRC-aided CCs for rate-$(\omega-1)/\omega$ CCs at short blocklength for the BI-AWGN channel, where the CC is either ZT or TB. We consider systematic, rate-$(\omega-1)/\omega$ convolutional encoders. The resulting concatenated codes are respectively called a CRC-ZTCC and a CRC-TBCC. We assume that the SLVD has a sufficiently large list size such that a codeword whose input sequence passes the CRC verification can always be found. Thus, the SLVD is an implementation of maximum-likelihood decoding. The FER is in fact the undetected error probability. Simulations show that at short blocklength, our rate-$(\omega-1)/\omega$ CRC-TBCCs still perform closely to the RCU bound. 

In \cite{Karimzadeh2020},  Karimzadeh and Vu considered designing the optimal CRC polynomial for multi-input CCs. In their framework, the information sequence is first divided into $(\omega-1)$ streams, one for each input rail, and they aim at designing an optimal CRC polynomial for each rail. Unlike their architecture, in this paper, the information sequence is first encoded with a single CRC polynomial and is then divided into $(\omega-1)$ streams. Simulation results show that our framework can yield better FER performance than that of Karimzadeh and Vu.

For rate-$(\omega-1)/\omega$ CCs, the SLVD on the \emph{primal trellis} requires high decoding complexity because of the $2^{\omega-1}$ outgoing branches at each node. SLVD implementation becomes complicated when there exist more than two outgoing branches per state. In order to simplify SLVD implementation and reduce complexity, we utilize the \emph{dual trellis} pioneered by Yamada \emph{et al}.\cite{Yamada1983}. The dual trellis expands the length of the primal trellis by a factor of $\omega$, while reducing the number of outgoing branches at each node from $2^{\omega-1}$ to at most two. %In \cite{Hartmann1976}, Hartmann \emph{et al.} derived the symbol-by-symbol decoding over the dual code that minimizes the symbol error probability. 
It is also worth noting that in \cite{Srinivasan2010}, Srinivasan \emph{et al.} derived the dual \emph{a posteriori} probability decoding of a high-rate CC.
%This achieves low decoding complexity by the SLVD on the corresponding dual trellis of the high-rate CC. 

The remainder of this paper is organized as follows. Section \ref{sec:SystematicAndDUalTrellis} reviews systematic encoding for $(\omega, \omega-1, v)$ convolutional codes and describes the dual trellis construction. Section \ref{sec: CRC-ZTCC} considers CRC-ZTCCs for rate-$(\omega-1)/\omega$ CCs. It addresses the zero-termination issue, presents DSO CRC design for high-rate ZTCCs, and shows CRC-ZTCC simulation results. Section \ref{sec: CRC-TBCC} considers CRC-TBCCs for rate-$(\omega-1)/\omega$ CCs. It addresses how to find the TB initial state over the dual trellis, describes DSO CRC design for TBCCs, and shows CRC-TBCC simulation results.  Section \ref{sec: conclusion} concludes the paper.

\textit{Notation}: Let $K$ and $N$ denote the information length and  blocklength in bits. Let $R = K/N$ denote the rate of a code with information length $K$ and blocklength $N$. A degree-$m$ CRC polynomial is of the form $p(x) = 1 + p_1x + \cdots + p_{m-1}x^{m-1} + x^m$, where $p_i\in\{0,1\}$, $i=1, 2, \dots, m-1$. For brevity, a CRC polynomial is represented in hexadecimal when its binary coefficients are written from the highest to lowest order. For instance, 0xD represents $x^3 + x^2 + 1$. The codewords are BPSK modulated. The SNR is defined as $\gamma_s \triangleq 10\log_{10}(A^2)$ (dB), where $A$ represents the BPSK amplitude and the noise follows a standard normal distribution.

% Yamada \emph{et al.}\cite{Yamada1983ANM} introduce techniques for using a dual-trellis with Viterbi decoding for high-rate convolutional codes (CCs), specifically rate-$(n-1)/n$ codes. This greatly decreases the decoding complexity compared to a normal trellis while maintaining the same error rate performance.

% Yang \emph{et al.} discuss the decoding performance of CC with designed distance-spectrum-optimal (DSO) cyclic redundancy checks (CRCs) using the serial list Viterbi decoding (SLVD). [other papers about CC-CRC...]

% This paper further develops the potential of using a convolutional code (CC) with a DSO CRC decoded using the SLVA over a dual-trellis for short-blocklength communication using both zero-terminated CCs (ZTCCs) and tail-biting CCs (TBCCs). Throughout this paper, $v$ represents the number of memory elements in the convolutional encoder and $m$ represents the degree of CRC. 

\section {Systematic Encoding and Dual Trellis}
\label{sec:SystematicAndDUalTrellis}
This section describes systematic encoding and introduces the dual trellis proposed by Yamada \emph{et al.} \cite{Yamada1983} for high-rate CCs generated with an $(\omega, \omega-1, v)$ convolutional encoder, where $v$ represents the overall constraint length. 

\subsection{Systematic Encoding}
We briefly follow \cite[Chapter 11]{LinCostello2004} in describing a systematic $(\omega, \omega-1, v)$ convolutional encoder. A systematic $(\omega, \omega-1, v)$ convolutional encoder can be represented by its parity check matrix
\begin{align}
    H(D) = [h^{(\omega-1)}(D), h^{(\omega-2)}(D), \dots, h^{(0)}(D)],
\end{align}
where each $h^{(i)}(D)$ is a polynomial of degree up to $v$ in delay element $D$ associated with the $i$-th code stream, i.e., 
\begin{align}
    h^{(i)}(D) = h_{v}^{(i)}D^v + h_{v-1}^{(i)}D^{v-1} + \cdots + h_0^{(i)},
\end{align}
where $h_j^{(i)}\in\{0, 1\}$. For convenience, we represent each $h^{(i)}(D)$ in octal form. For instance, $H(D)=[D^3+D^2+D+1, D^3+D^2+1, D^3+D+1]$ can be concisely written as $H=(17, 15, 13)$. Define $\bm{h}^{(i)} \triangleq [h_v^{(i)}, h_{v-1}^{(i)}, \dots, h_0^{(i)}]$, $i=0, 1, \dots, \omega-1$. The systematic encoding matrix $G(D)$ associated with $H(D)$ is given by
\begin{align}
    G(D) = \begin{bmatrix}
    \frac{h^{(1)}(D)}{h^{(0)}(D)} & 1 & 0 & \cdots & 0\\
    \frac{h^{(2)}(D)}{h^{(0)}(D)} & 0 & 1 & \cdots & 0\\
    \vdots & \vdots & \vdots & \ddots & \vdots \\
    \frac{h^{(\omega-1)}(D)}{h^{(0)}(D)} & 0 & 0 & \cdots & 1
    \end{bmatrix}.
\end{align}
The first output bit is a coded bit and the remaining output bits are a direct copy of the corresponding input bits.

\subsection{Dual Trellis}
The primal trellis associated with a rate-$(\omega-1)/\omega$ ZTCC has $2^{\omega-1}$ outgoing branches per state. Performing SLVD over the primal trellis when $\omega > 2$ is highly complex. In \cite{Yang2022}, the low decoding complexity of SLVD for rate-$1/\omega$ convolutional codes relies on the fact that only $2$ outgoing branches are associated with each state. In order to efficiently perform SLVD, we consider the dual trellis proposed by Yamada \emph{et al.} \cite{Yamada1983}. 

We briefly explain the dual trellis construction for parity check matrix $H(D) = [h^{(\omega-1)}(D), h^{(\omega-2)}(D), \dots, h^{(0)}(D)]$. First, we define the maximum instant response order $\lambda$ as
\begin{align}
    \lambda \triangleq \max\{j\in\Brace{0,1,\dots, \omega-1}:  h_0^{(j)}=1\}.
\end{align}
The state of the dual trellis is represented by the partial sums of $(v+1)$ adders in the observer canonical form of $H(D)$. At time index $j$, $j=0,1,\dots, \omega-1$, the state is given by
\begin{align}
    \bms^{(j)} = [s_v^{(j)}, s_{v-1}^{(j)},  \dots, s_0^{(j)}].
\end{align}
Next, we show how the state $\bms^{(j)}$ evolves in terms of the output bits $\bmy_k = [y_k^{(0)}, y_k^{(1)}, \dots, y_k^{(\omega-1)}]$, $k=1,2,\dots, N/\omega$, so that a dual trellis can be established.

\textit{Dual trellis construction for $\bmy_k = [y_k^{(0)}, y_k^{(1)}, \dots, y_k^{(\omega-1)}]$}:
\begin{itemize}
    \item[1)] At time $j=0$, $\bms^{(0)}=[0, s_{v-1}^{(j)}, s_{v-2}^{(j)}, \dots, s_{0}^{(j)}]$, where $s_i^{(0)}\in\{0,1\}$. Namely, only $2^v$ states exist at $j=0$.
  \item[2)] At time $j$, $j<\omega-1$, draw branches from each state $\bms^{(j)}$ to state $\bms^{(j+1)}$ by
  \begin{align}
      \bms^{(j+1)} = \bms^{(j)} + y_k^{(j)}\bmh^{(j)},\quad  y_k^{(j)}=0,1. \label{eq: transition}
  \end{align}
  \item[3)] At time $j=\omega-1$, draw branches from each state $\bms^{(\omega-1)}$ to state $\bms^{(\omega)}$ by
  \begin{align}
      \bms^{(\omega)} = \Big(\bms^{(\omega-1)} + y_k^{(\omega-1)}\bmh^{(\omega-1)}\Big)^r,\   y_k^{(\omega-1)}=0,1,
  \end{align}
  where $(a_v, a_{v-1}, \dots,a_{1}, a_0)^r = (0, a_v, a_{v-1}, \dots, a_1)$.
  \item[4)] For time $j = \lambda$, draw a branch from each state $\bms^{(\lambda)}$ according to \eqref{eq: transition} only for $y_k^{(\lambda)} = s_0^{(\lambda)}$.
\end{itemize}
After repeating the above construction for each $\bmy_k$, $k = 1, 2,\dots, N/\omega$, we obtain the dual trellis associated with the $(\omega, \omega-1, v)$ convolutional code. Since the primal trellis is of length $N/\omega$, whereas the dual trellis is of length $N$, the dual trellis can be thought of as expanding the primal trellis length by a factor of $\omega$, while reducing the number of outgoing branches per state from $2^{\omega-1}$ to less than or equal to $2$.

% \textcolor{red}{(Introduce the dual trellis here)}
\section{ZTCC with DSO CRC via Dual Trellis SLVD}\label{sec: CRC-ZTCC}

This section considers CRC-ZTCCs for rate-$(\omega-1)/\omega$ CCs. Section \ref{subsec: zero termination} presents a zero termination method over the dual trellis. Section \ref{subsec: DSO CRC ZTCC} describes our DSO CRC polynomial search procedure. Finally, Section \ref{subsec: results ZTCC} presents simulation results of the CRC-ZTCC compared with the RCU bound. As a case study, this paper mainly focuses on the rate-$3/4$ systematic feedback convolutional codes in \cite[Table 12.1(e)]{LinCostello2004}.

%The dual trellis reduces the number of out going branches by using the parity check matrix instead of the coder generator polynomial matrix, as well as by using adders in the syndrome former as memory elements. For any rate-$(\omega-1)/\omega$ convolutional code, the number of adders is always one greater than the number of memory elements $v$. Therefore, the number of "states" in a dual trellis is doubled compared to a primal trellis. 

%A common approach for achieving ZTCC is to append $v$ bits of zeros at the end of the original message. The zero-padding method aims to make all memory elements zero, which corresponds to the all-zero state in a primal trellis. However, for dual trellis, because we are using adders as memory elements, the zero-padding method has the issue that it cannot essentially bring all the adder values back to zero and reach the all-zero state. 

%Since there are $2^v$ total states and $2^\omega$ branches from each state in the primal trellis, we can go from any state back to the all-zero state in $2^{v-\omega}$ transitions. That is equivalent to $\omega 2^{v-\omega}$ overhead bits in the dual trellis needed to bring the sequence back to the all-zero state. For the rate-$3/4$ ZTCC with feedback encoder $(33, 25, 37, 31)$, we generate a map that provides the bits needed for each state to fulfill the condition of ZT. These overhead bits greatly reduce the performance of the list decoder and they also decrease the channel coding rate of the codes.

\begin{table}[t] 
\centering
\caption{DSO CRC Polynomials for Rate-$3/4$ ZTCC at Blocklength $N=128$ Generated by $H=(33, 25, 37, 31)$ With $v=4$, by $H=(47, 73, 57, 75)$ With $v=5$, and by $H=(107, 135, 133, 141)$ With $v=6$}
\label{tab:ZT_codes_with_CRC}
\begin{tabular}{P{0.3cm} P{0.2cm} P{0.6cm} P{1.4cm} P{1.4cm} P{1.4cm}}
\hline
\clineB{1-6}{1.2}
& \\[\dimexpr-\normalbaselineskip+2pt]
$K$ & $m$ & $R$ & $v=4$ CRC & $v=5$ CRC & $v=6$ CRC \\ 
 \hline 
%  \\[\dimexpr-\normalbaselineskip+2pt]
$87$ &$3$ & $0.680$ & 0x9 & 0x9 & 0xB \\
%  & \\[\dimexpr-\normalbaselineskip+2pt]
$86$ &$4$ & $0.672$ & 0x1B & 0x15 & 0x1D \\
%  & \\[\dimexpr-\normalbaselineskip+2pt]
$85$ &$5$ & $0.664$ & 0x25  & 0x25 & 0x25 \\
%   & \\[\dimexpr-\normalbaselineskip+2pt]
$84$ &$6$ & $0.656$ & 0x4D & 0x7B & 0x6F \\
%   & \\[\dimexpr-\normalbaselineskip+2pt]
$83$ &$7$ & $0.648$ & 0xF3 & 0xED & 0x97 \\
%  & \\[\dimexpr-\normalbaselineskip+2pt]
$82$ &$8$ & $0.641$ & 0x1E9 & 0x1B7 & 0x1B5 \\ 
%  & \\[\dimexpr-\normalbaselineskip+2pt]
$81$ &$9$ & $0.633$ & 0x31B & 0x3F1 & 0x2F1 \\
%  & \\[\dimexpr-\normalbaselineskip+2pt]
$80$ &$10$ & $0.625$ & 0x5C9 & 0x66F & 0x59F \\
%  & \\[\dimexpr-\normalbaselineskip+2pt]
 \hline
\clineB{1-6}{1.2}
\end{tabular}
\end{table}

\subsection{Zero Termination of Dual Trellis}\label{subsec: zero termination}

For an $(\omega, \omega-1, v)$ CC, zero termination over the dual trellis requires at most $\omega\lceil v/(\omega-1) \rceil$ steps. In our implementation, a breadth-first search identifies the zero-termination input and output bit patterns that provide a trajectory from each possible state $\bms$ to the zero state. The input and output bit patterns have lengths $(\omega-1)\lceil v/(\omega-1) \rceil$ and $\omega\lceil v/(\omega-1) \rceil$ respectively.

\subsection{Design of DSO CRCs for High-Rate ZTCCs}\label{subsec: DSO CRC ZTCC}

In general, a DSO CRC polynomial provides the optimal distance spectrum which minimizes the union bound on the FER at a specified SNR \cite{Yang2022}. In this paper, we focus on the low FER regime. Thus, the DSO CRC polynomials identified in this paper simply maximize the minimum distance of the concatenated code. Examples in \cite{Yang2022} indicate that DSO CRC polynomials designed in this way can provide optimal or near-optimal performance for a wide range of SNRs.

The design procedure of the DSO CRC polynomial for high-rate ZTCCs essentially follows from the DSO CRC design algorithm for low target error probability in \cite{Yang2022}. The first step is to collect the \emph{irreducible error events} (IEEs), which are ZT paths on the trellis that deviate from the zero state once and rejoin it once. IEEs with a very large output Hamming weight do not affect the choice of optimal CRCs.  In order to reduce the runtime of the CRC optimization algorithm, IEEs with output Hamming weight larger than a threshold $\tilde{d}-1$ are not considered. Dynamic programming constructs all ZT paths of length equal to $N/\omega$ and output weight less than $\tilde{d}$. Finally, we use the resulting set of ZT paths to identify the degree-$m$ DSO CRC polynomial for the rate-$(\omega-1)/\omega$ CC.

\begin{figure}[t]
\centering
\includegraphics[width=3in]{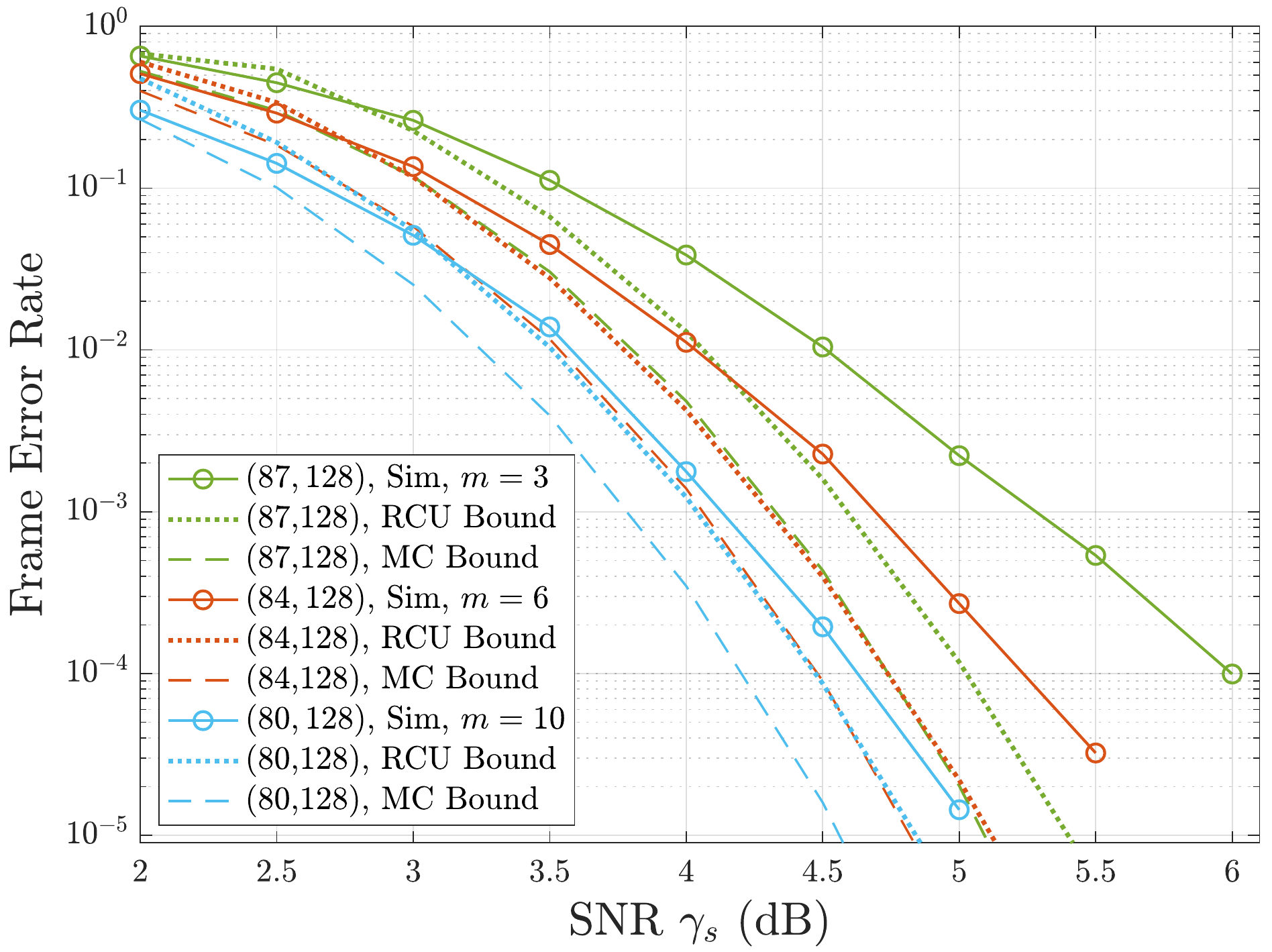}
\caption{FER vs. SNR for various CRC-ZTCCs. The ZTCC is generated with the $(4,3,6)$ encoder $H=(107, 135, 133, 141)$. The DSO CRC polynomials of degrees $3, 6,$ and $10$ are 0xB, 0x6F, and 0x59F, respectively. Values in parenthesis denote information length $K$ and blocklength $N$, respectively. }
\label{Fig:ZT_FER_SNR_v6_N128}
\end{figure}

Table \ref{tab:ZT_codes_with_CRC} presents the DSO CRC polynomials for ZTCCs generated with $H=(33, 25, 37, 31)$, $H=(47, 73, 57, 75)$, and $H=(107, 135, 133, 141)$. The design assumes a fixed blocklength $N = 128$ bits. Due to the overhead caused by the CRC bits and by zero termination, the rates of CRC-ZTCCs are less than $3/4$. Specifically, for a given information length $K$, CRC degree $m$ and an $(\omega, \omega-1, v)$ encoder, the blocklength $N$ for a CRC-ZTCC is given by
$
    N = \left (K+m + (\omega-1) \left \lceil \frac{v}{\omega-1} \right \rceil \right ) \frac{\omega}{\omega-1} \label{eq: blocklength ZTCC}
$, giving
$
    R= \frac{K}{N} = \frac{\omega-1}{\omega}\frac{K}{K+m+(\omega-1)\lceil \frac{v}{\omega-1} \rceil }
$. We see from \eqref{eq: blocklength ZTCC} that the $(\omega, \omega-1, v)$ convolutional encoder can accept any CRC degree $m$ as long as $K+m$ is divisible by $(\omega-1)$.

% \textcolor{red}{add m=8/10 cases to Table I?}

\subsection{Results and Comparison with RCU Bound}\label{subsec: results ZTCC}

Fig. \ref{Fig:ZT_FER_SNR_v6_N128} shows the performance of CRC-ZTCCs with increasing CRC degrees $3, 6$ and $10$ and a fixed blocklength $N=128$ bits. We see that at the target FER of $10^{-4}$, increasing the CRC degree reduces the gap to the RCU bound. With $m=10$ and $v=6$, the CRC-ZTCC approaches the RCU bound within $0.25$ dB.

%where the second term in the numerator is the number of CRC bits and the third term is the number of zero-termination bits. Thus, given the same blocklength $N$ and ZTCC encoder, the channel coding rate decreases as the CRC degree increases, but the overhead bits have less influence on the channel coding rate. This observation is reflected in Fig.\ref{Fig:ZT_FER_SNR_v6_N128} that the simulation result of CRC-ZTCC with CRC (0x31B) has a gap of $1.3$ dB with the RCU bound at $\text{FER} = 10^{-4}$, which is the least among all three CRCs.

\begin{figure}[t]
\centering
\includegraphics[width=3.1in]{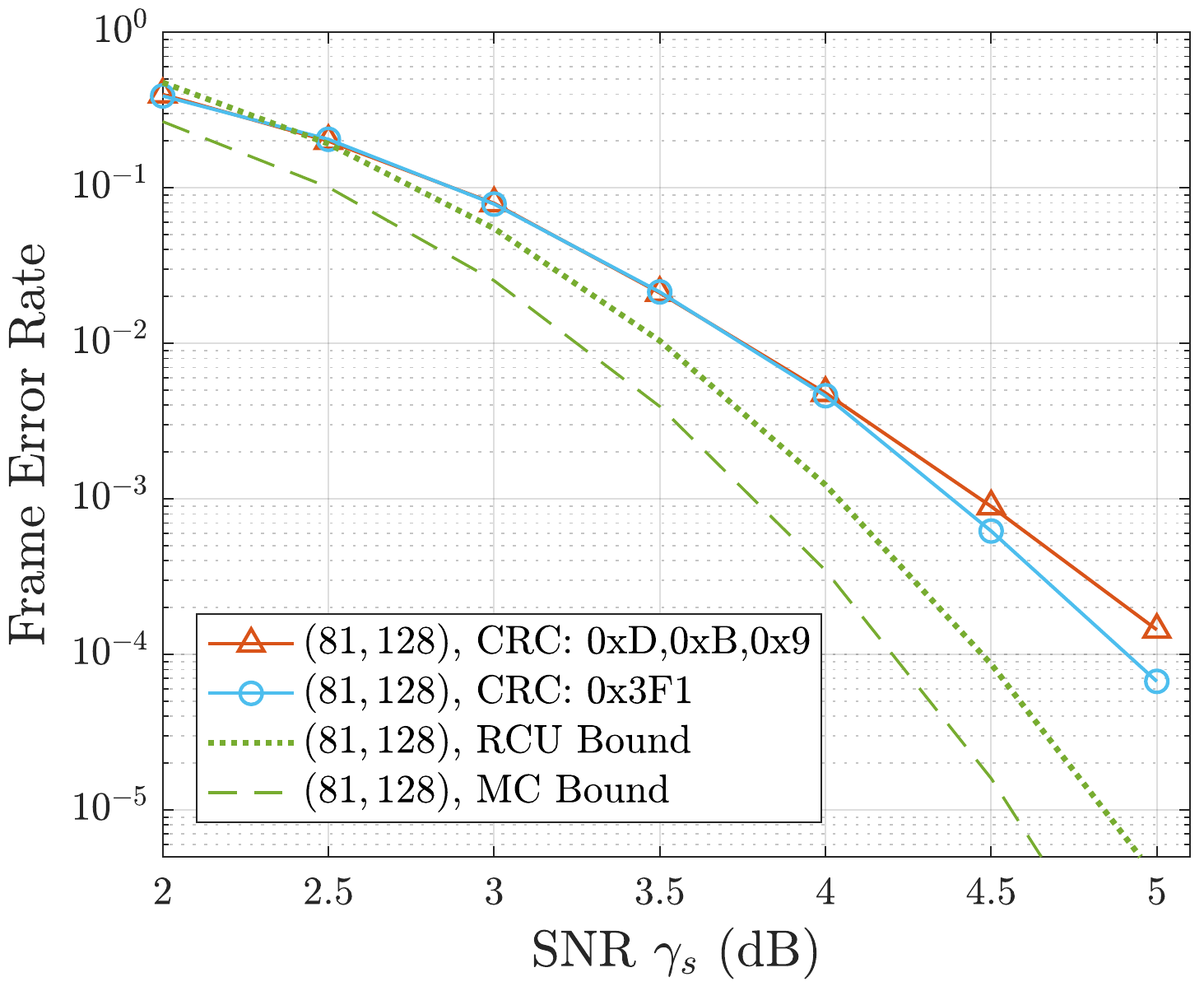}
\caption{FER vs. SNR for $v = 5$ CRC-ZTCCs designed under Karimzadeh \emph{et al.}'s scheme \cite{Karimzadeh2020} and our scheme. Both CRC-ZTCCs have information length $K=81$ and blocklength $N=128$.}
\label{Fig:ZT_FER_compare}
\end{figure}

In \cite{Karimzadeh2020}, Karimzadeh \emph{et al.} considered designing optimal CRC polynomials for each input rail of a multi-input CC. In their setup, an information sequence for an $(\omega, \omega-1, v)$ encoder needs to be split into $(\omega-1)$ subsequences before CRC encoding. In contrast, the entire information sequence in our framework is encoded with a single CRC polynomial. Then the resulting sequence is evenly divided into $(\omega-1)$ subsequences, one for each rail. To compare the performance between these two schemes, we design three degree-$3$ optimal CRC polynomials, one for each rail, for ZTCC with $H=(47, 73, 57, 75)$. The three CRC polynomials jointly maximize the minimum distance of the CRC-ZTCC. For the single-CRC design, we use the single degree-$9$ DSO CRC polynomial for the same encoder from Table \ref{tab:ZT_codes_with_CRC}. Both CRC-ZTCCs have an information length $K=81$ and blocklength $N = 128$. Fig. \ref{Fig:ZT_FER_compare} shows the performance comparison between these two codes, showing that a single degree-$9$ DSO CRC polynomial outperforms three degree-$3$ DSO CRC polynomials, one for each rail. This suggests that a single DSO CRC polynomial may suffice to provide superior protection for each input rail. The decoding complexity is similar regardless of the CRC scheme.

%Fig.\ref{Fig:ZT_FER_compare} demonstrates a performance comparison between the degree-$9$ DSO CRC designed according to \ref{subsec: DSO CRC ZTCC} and three degree-$3$ CRCs proposed by Karimzadeh and Vu \cite{Vu2020}. The three CRCs each append at the end of one input rails, resulting in an overall degree-$9$ CRC. The performance of the two CRC designs is very similar till $\gamma_s = 4$, where a considerate gap starts growing between the curves. At $\text{FER}=10^{-3}$, the gap is about $0.1$ dB and it doubles at $\text{FER}=10^{-4}$. Thus we use DSO CRCs for the rest of this paper.

\section{TBCC with DSO CRC and Dual Trellis SLVD}\label{sec: CRC-TBCC}
The performance of CRC-aided list decoding of ZTCCs relative to the RCU bound is constrained by the termination bits appended to the end of the original message, which are required to bring the trellis back to the all-zero state. TBCCs avoid this overhead by replacing the zero termination condition with the TB condition, which states that the final state of the trellis is the same as the initial state of the trellis \cite{Ma1986}.

In this section, we apply the SLVD to CRC-TBCCs over the dual trellis. We will discuss how to determine the initial state for the TBCC to ensure that the TB condition is met and demonstrate designs of DSO CRCs for rate $(\omega-1)/\omega$ TB codes. Decoding complexity and performance are analyzed at the end of this section. 

\subsection{List Decoding for TBCC with CRC}

There are two primary differences in the development and analysis of list decoding between ZTCCs, as described in Section \ref{sec: CRC-ZTCC}, and TBCCs. One difference is that since the ZT condition is replaced with the TB condition, the encoder must determine the initial trellis state so that the TB condition is satisfied. The other difference is that SLVD on the dual trellis must be adapted to handle the TB condition.

To satisfy the TB condition, encoding is attempted from every initial state to identify the initial state that satisfies the TB condition. This is required because our recursive encoder cannot simply achieve the TB condition by setting the initial encoder memory to be the final bits of the information sequence.

% As \cite{Yang2022} stated, the TB condition can be achieved by setting the initial encoder memory to be the final bits of the information sequence for feedforward CCs when encoding the message, which is not feasible for feedback cases. Thus, we perform encoding from every possible initial state and check if the codeword terminates in the same state. 

% At the decoder end, unlike list decoding for ZTCC, where we start every trackback from the all-zero state, it's crucial to determine the right terminating state to start from so that we traverse the paths in ascending path metrics for the TBCC list decoder. 

\begin{figure}[t]
\centering
\includegraphics[width=17.5pc]{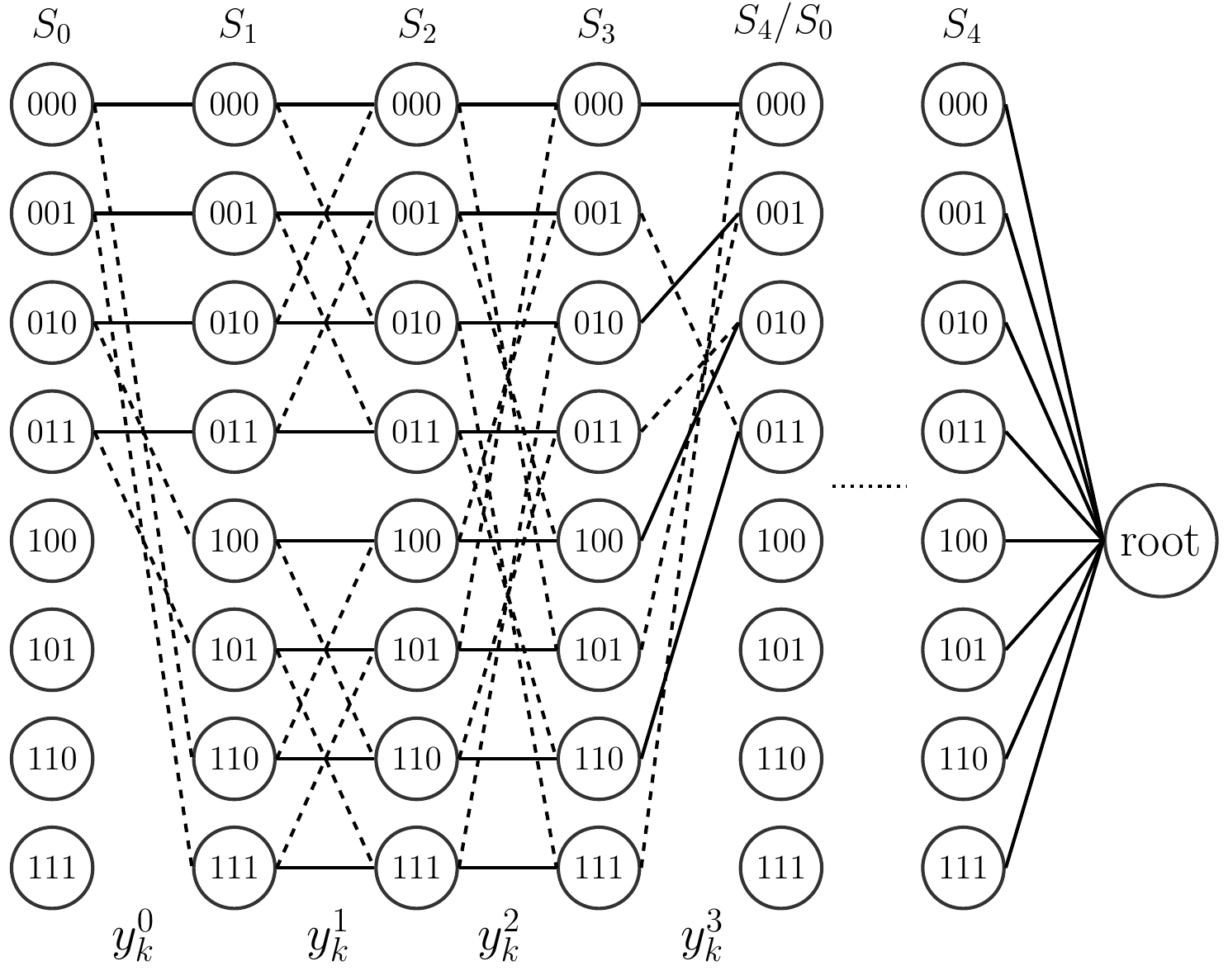}
\caption{Dual trellis diagram for rate-$3/4$ TBCC with a root node at the end for encoder $H=(2,5,7,6)$ with $v=2$. Solid lines represent $0$ paths and dashed lines represent $1$ paths.}
\label{Fig:root}
\end{figure}

To adapt SLVD on the dual trellis to handle the TB condition, we propose an efficient way to keep track of the path metrics and find the next path with minimum metric through an additional root node as shown in Fig. \ref{Fig:root}. The root node connects to all terminating states after forward traversing the dual trellis. The Hamming distance of the branch metric for the branch connecting any state to this root node is zero. This additional root node allows the trellis to end in a single state, so that the basic SLVD approach for a ZTCC may be applied. During SLVD, if the current path does not pass either the CRC or TB check, the minimum value among all remaining path metrics will be selected as the next path to check.

% \subsection{Finding Initial State for TBCC on the Dual Trellis}

% \textcolor{red}{(To be written)}

\subsection{Design of DSO CRCs for High-Rate TBCCs}

The design of DSO CRCs for high-rate TBCCs follows the two-phase design algorithm shown in \cite{Yang2022}. This algorithm is briefly explained below.

Consider a TB trellis $T = (V, E, \A)$ of length $n$, where $\A$ denotes the set of output alphabet, $V$ denotes the set of states, and $E$ denotes the set of edges described in an ordered triple $(s, a, s')$ with $s, s'\in V$ and $a\in\A$ \cite{Koetter2003}. Assume $|V| = 2^v$ and let $V_0 = \{0, 1, \dots, 2^v-1\}$. Define the set of IEEs at state $\sigma\in V$ as
\begin{align}
    \IEE(\sigma)\triangleq\bigcup_{l=1,2,\dots,n}\overline{\IEE}(\sigma,l),
\end{align}
where
\begin{align}
    \overline\IEE(\sigma,l)\triangleq&\{(\bm{s},\bm{a})\in V_0^{l+1}\times \A^{l}: s_0=s_l=\sigma;\notag\\
  &\forall j, 0<j<l,\ s_{j}\notin\{0, 1,\dots,\sigma\}\}.
\end{align}
The IEEs at state $\sigma$ can be thought of as ``building blocks'' for an arbitrarily long TB path that starts and ends at the same state $\sigma$.

The first phase is called the collection phase, during which the algorithm collects $\IEE(\sigma)$ with output Hamming weight less than the threshold $\tilde{d}$ over a sufficiently long TB trellis. The second phase is called the search phase, during which the algorithm first reconstructs all TB paths of length $N/\omega$ and output weight less than $\tilde{d}$ via concatenation of the IEEs and circular shifting of the resulting path. Then, using these TB paths, the algorithm searches for the degree-$m$ DSO CRC polynomial by maximizing the minimum distance of the undetected TB path.

Table \ref{tab:TB_codes_with_CRC} presents the DSO CRC polynomials for TBCCs generated with $H=(33, 25, 37, 31)$, $H=(47, 73, 57, 75)$, and $H=(107, 135, 133, 141)$. The design assumes a fixed blocklength $N = 128$. TB encoding avoids the rate loss caused by the overhead of the zero termination. Specifically, for a given information length $K$, CRC degree $m$ and an $(\omega, \omega-1, v)$ encoder, the blocklength $N$ for a CRC-TBCC is given by
$
    N = \left( {K+m} \right ) \frac{\omega}{\omega-1} \label{eq: blocklength}
$, giving
$
    R = \frac{K}{N} = \frac{\omega-1}{\omega}\frac{K}{K+m}
$.

\begin{table}[t]
\centering
\caption{DSO CRC Polynomials for rate-$3/4$ TBCC at Blocklength $N=128$ Generated by $H=(33, 25, 37, 31)$ With $v=4$, by $H=(47, 73, 57, 75)$ With $v=5$, and by $H=(107, 135, 133, 141)$ With $v=6$}
\label{tab:TB_codes_with_CRC}
\begin{tabular}{P{0.3cm} P{0.2cm} P{0.6cm} P{1.4cm} P{1.4cm} P{1.4cm}}
\hline
\clineB{1-6}{1.2}
& \\[\dimexpr-\normalbaselineskip+2pt]
$K$ & $m$ & $R$ & $v=4$ CRC & $v=5$ CRC & $v=6$ CRC \\ 
 \hline 
%  \\[\dimexpr-\normalbaselineskip+2pt]
$93$ &$3$ & $0.727$ & 0x9 & 0x9 & 0xB \\
%  & \\[\dimexpr-\normalbaselineskip+2pt]
$92$ &$4$ & $0.719$ & 0x1B & 0x1D & 0x17 \\
%  & \\[\dimexpr-\normalbaselineskip+2pt]
$91$ &$5$ & $0.711$ & 0x25  & 0x3B & 0x33 \\
%   & \\[\dimexpr-\normalbaselineskip+2pt]
$90$ &$6$ & $0.703$ & 0x7D & 0x4F & 0x41 \\
%   & \\[\dimexpr-\normalbaselineskip+2pt]
$89$ &$7$ & $0.695$ & 0xF9 & 0xD1 & 0xBD \\
%  & \\[\dimexpr-\normalbaselineskip+2pt]
$88$ &$8$ & $0.688$ & 0x1CF & 0x173 & 0x111 \\ 
%  & \\[\dimexpr-\normalbaselineskip+2pt]
$87$ &$9$ & $0.680$ & 0x38F & 0x3BF & 0x333 \\
%  & \\[\dimexpr-\normalbaselineskip+2pt]
$86$ &$10$ & $0.672$ & 0x73F & 0x697 & 0x723 \\
%  & \\[\dimexpr-\normalbaselineskip+2pt]
 \hline
\clineB{1-6}{1.2}
\end{tabular}
\end{table}

\subsection{Complexity Analysis}\label{subsec: complexity}
In \cite{Yang2022}, the authors provided the complexity expression for SLVD of CRC-ZTCCs and CRC-TBCCs, where the convolutional encoder is of rate $1/\omega$. Observe that the dual trellis has no more than $2$ outgoing branches per state, similar to the trellis of a rate-$1/\omega$ CC. Thus, we directly apply their complexity expression to the SLVD over the dual trellis.

\begin{figure}[t]
\centering
\includegraphics[width=3.1in]{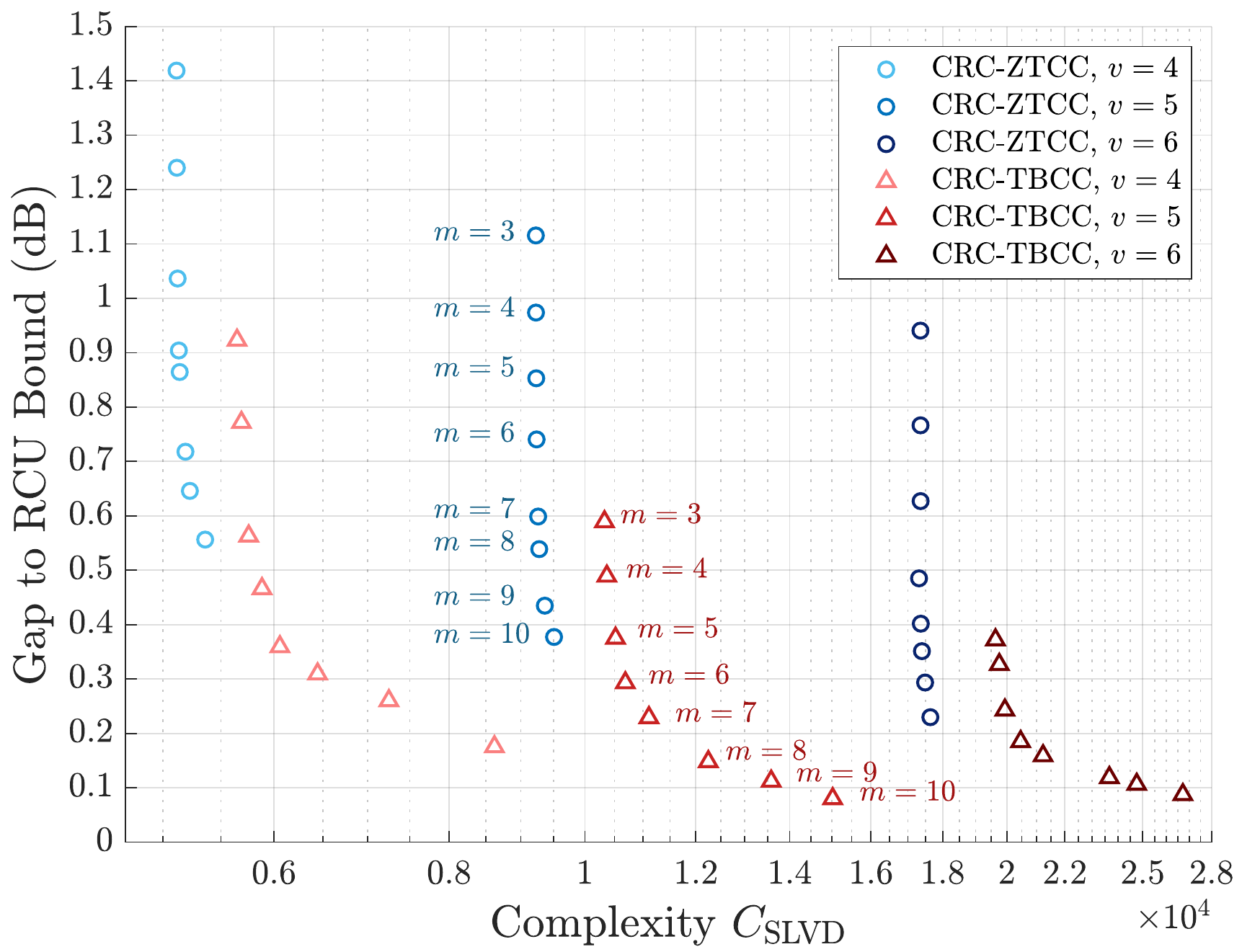}
\caption{The SNR gap to the RCU bound vs. the average complexity of SLVD of CRC-ZTCC codes in Table I and CRC-TBCC codes in Table II for target FER of $10^{-4}$. Each color represents a specific CRC-aided CC shown in the tables. Markers from top to bottom with the same color correspond to DSO CRC polynomials with $m$ = $3$, $\dots$, $10$.}
\label{Fig:SNR_Complexity}
\end{figure}

As noted in \cite{Yang2022}, the overall average complexity of the SLVD can be decomposed into three components:
\begin{align}
    C_{\SLVD} = C_{\SSV} + C_{\trace} + C_{\List},
\end{align}
where $C_{\SSV}$ denotes the complexity of a standard soft Viterbi (SSV), $C_{\trace}$ denotes the complexity of the \emph{additional} traceback operations required by SLVD, and $C_{\List}$ denotes the average complexity of inserting new elements to maintain an ordered list of path metric differences. 

$C_{\SSV}$ is the complexity of add-compare-select (ACS) operations and the initial traceback operation. For CRC-ZTCCs,
\begin{align}
    C_{\SSV}&=(2^{v+1}-2)+1.5(2^{v+1}-2)+1.5(K+m-v)2^{v+1}\notag\\
  &\phantom{=}+c_1[2(K+m+v)+1.5(K+m)]. \label{eq: C_SSV}
\end{align}
For CRC-TBCCs, this quantity is given by
\begin{align}
    C_{\SSV}&=1.5(K+m)2^{v+1}+2^{v}+3.5c_1(K+m). \label{eq: C_SSV for CRC-TBCC}
\end{align}
The second component $C_{\trace}$ for CRC-ZTCC is given by
\begin{align}
    C_{\trace} = c_1(\E[L]-1)[2(K+m+\nu)+1.5(K+m)]. \label{eq: C_trace}
\end{align}
For CRC-TBCCs, $C_{\trace}$ is given by
\begin{align}
    C_{\trace} = 3.5c_1(\E[L]-1)(K+m). \label{eq: C_trace for CRC-TBCC}
\end{align}
The third component, which is the same for ZT and TB, is
\begin{align}
    C_{\List} = c_2\E[I]\log(\E[I]), \label{eq: C_list}
\end{align}
where $\E[I]$ is the expected number of insertions to maintain the sorted list of path metric differences. For CRC-ZTCCs,
\begin{align}
  \E[I]&\le (K+m)\E[L], \label{eq: EL ZTCC}
\end{align}
and for CRC-TBCCs,
\begin{align}
  \E[I]&\le (K+m)\E[L]+2^{v}-1. \label{eq: EL TBCC}
\end{align}
In the above expressions, $c_1$ and $c_2$ are two computer-specific constants that characterize implementation-specific differences in the implemented complexity of traceback and list insertion (respectively) as compared to the ACS operations of Viterbi decoding. In this paper, we assume that $c_1 = c_2 = 1$ and use \eqref{eq: EL ZTCC} and \eqref{eq: EL TBCC} to estimate $\E[I]$ for CRC-ZTCCs and CRC-TBCCs, respectively.

% \begin{align}
%     \text{total \# edges} = 2^{\omega-1}\frac{2^{(\omega-1)\lceil\frac{v}{\omega-1} \rceil } - 1}{2^{\omega-1}-1}
% \end{align}

% The term $2^{\omega}$ is resulted from the branches connecting to the root node. We include this in the traceback matrix to simplify the process of transiting between the ending states. 

\begin{figure}[t]
\centering
\includegraphics[width=3.1in]{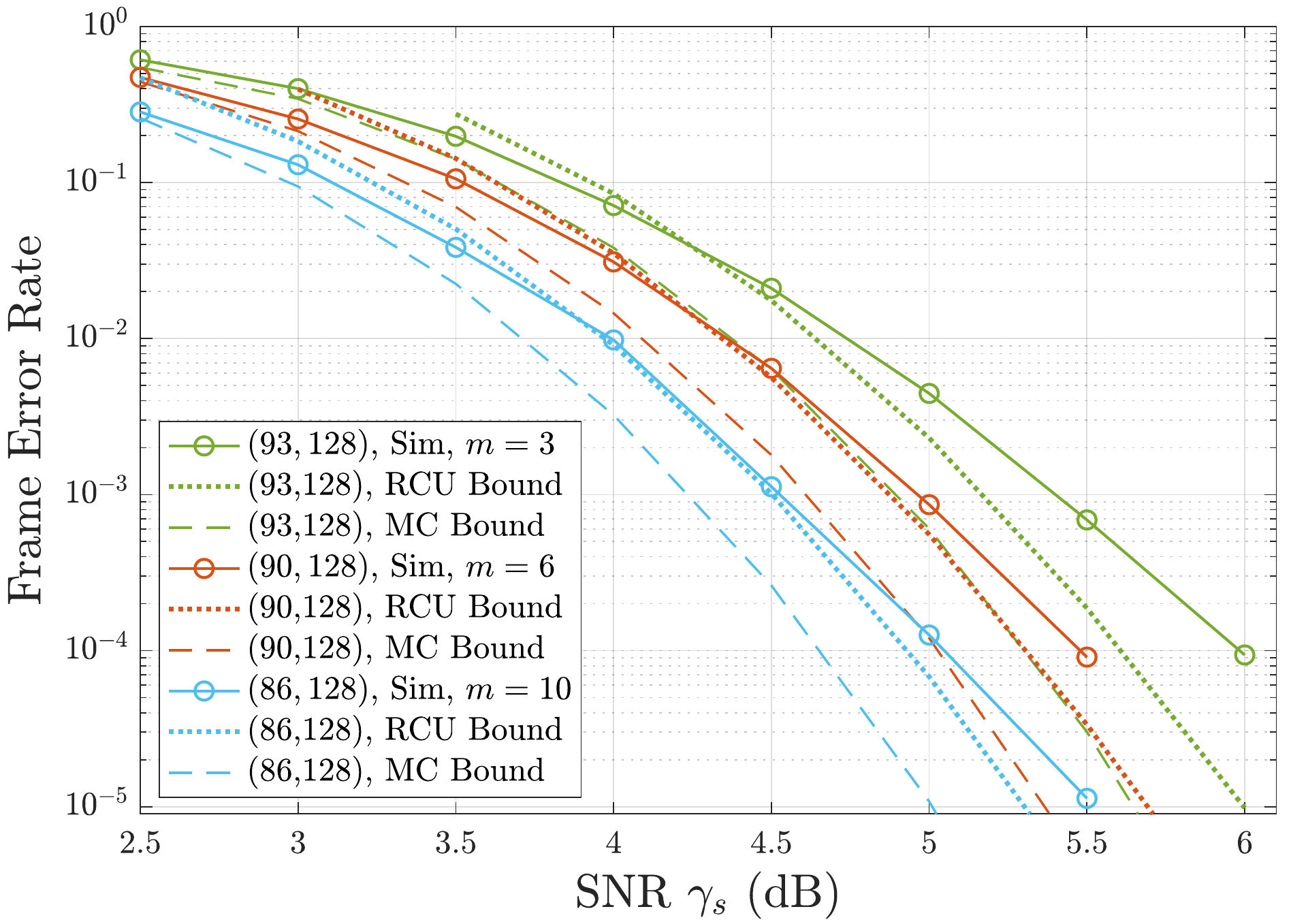}
\caption{FER vs. SNR for various CRC-TBCCs. The TBCC is generated with the $(4,3,6)$ encoder $H=(107, 135, 133, 141)$. The DSO CRC polynomials of degrees $3, 6,$ and $10$ are 0xB, 0x41, and 0x723, respectively. Values in parenthesis denote information length $K$ and blocklength $N$, respectively. }
\label{Fig:TB_FER_SNR_v6_N128}
\end{figure}

\begin{figure}[t]
\centering
\includegraphics[width=3.1in]{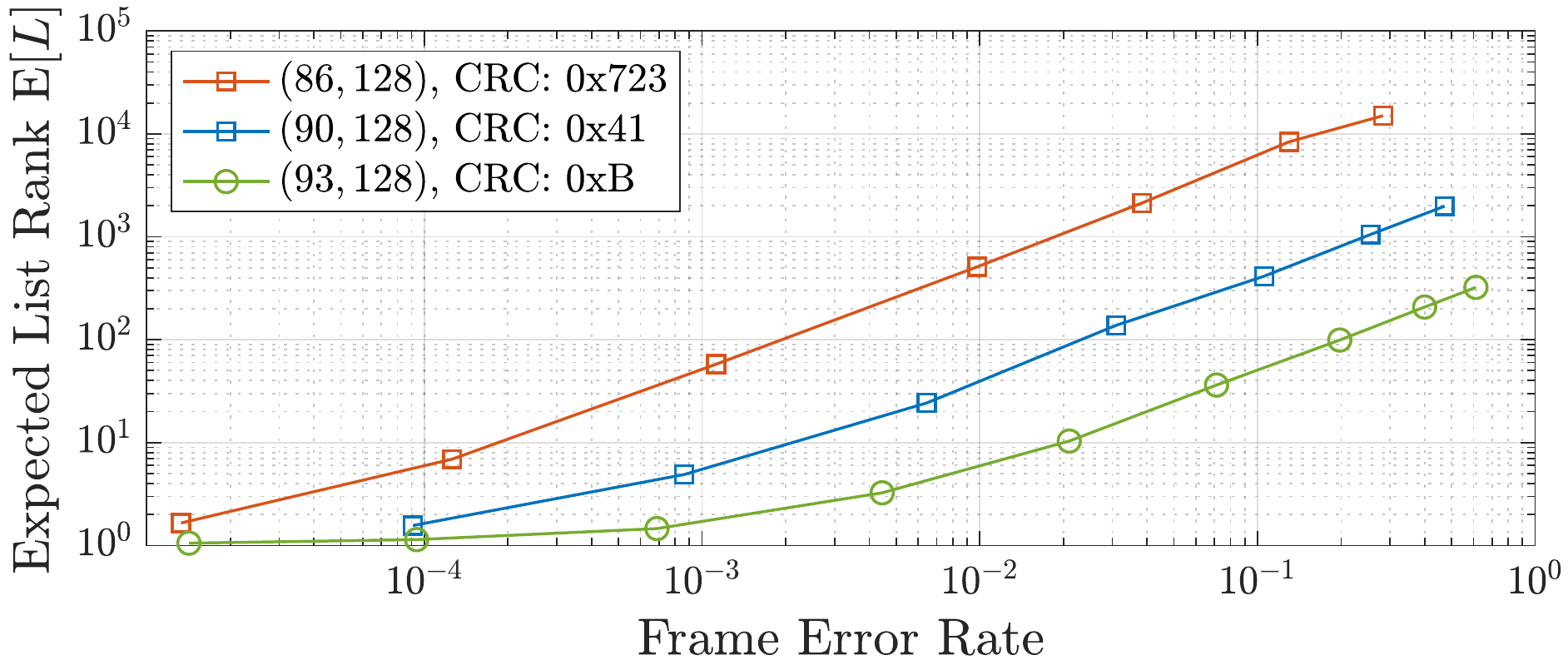}
\caption{Expected list rank $\E[L]$ vs. FER for SLVD of CRC-TBCC with blocklength $N=128$. The TBCC is generated with $H = (107, 135, 133, 141)$. The DSO CRC polynomials of degrees $3, 6$ and $10$ are 0xB, 0x41, and 0x723.}
\label{Fig:TB_EL_v5_N128}
\end{figure}

% Fig. \ref{Fig:SNR_Complexity} shows the trade-off between the SNR gap to the RCU bound and the average decoding complexity at the target FER $10^{-4}$. The average decoding complexity of SLVD is evaluated according to the aforementioned expressions. We see that for a fixed $v$ (ZT or TB), increasing the CRC degree $m$ significantly reduces the gap to the RCU bound, at the cost of a small increase in complexity. The minimum gap of $0.08$ dB is achieved by the CRC-TBCC with $v = 5$ and $m = 10$. However, for the same CRC degree $m$, increasing the overall constraint length $v$ dramatically increases the complexity, while achieving a minimal reduction in the SNR gap to the RCU bound.

%Fig. \ref{Fig:SNR_Complexity} shows the trade-off between the SNR gap to the RCU bound and the average decoding complexity for a target frame error rate $\text{FER}  = 10^{-4}$. For both ZTCCs and TBCCs, as the degree of DSO CRC $m$ increases, the decoding complexity $C_{\SLVD}$ grows slightly while the gap to RCU bound is reduced greatly. The reduction of gap to RCU bound is more significant for ZTCCs than TBCCs. For ZTCCs, the number of memory elements $\nu$ doesn't have an outstanding effect on the SNR gap reduction. But such reduction is considerable for TBCCs with different $\nu$. Given the same encoder polynomial and blocklength, CRC-TBCC outperforms CRC-ZTCC by a factor of $2$ for all CRC degrees. For both CRC-ZTCC and CRC-TBCC, the gap to RCU bound diminishes with a factor of around $1.3$ at the cost of nearly doubling the complexity with an additional memory element.  

\subsection{Results, Analysis, and Expected List Rank of SLVD}

Fig. \ref{Fig:SNR_Complexity} shows the trade-off between the SNR gap to the RCU bound and the average decoding complexity at the target FER $10^{-4}$ for CRC-ZTCCs designed in Table \ref{tab:ZT_codes_with_CRC} and CRC-TBCCs designed in Table \ref{tab:TB_codes_with_CRC}. The average decoding complexity of SLVD is evaluated according to the expressions in Sec. \ref{subsec: complexity}. We see that for a fixed $v$ (ZT or TB), increasing the CRC degree $m$ significantly reduces the gap to the RCU bound, at the cost of a small increase in complexity. The minimum gap of $0.08$ dB is achieved by the CRC-TBCC with $v = 5$ and $m = 10$. However, for the same CRC degree $m$, increasing the overall constraint length $v$ dramatically increases the complexity, while achieving a minimal reduction in the SNR gap to the RCU bound.

Fig. \ref{Fig:TB_FER_SNR_v6_N128} shows the FER vs. SNR for three CRC-TBCCs at blocklength $N = 128$. At the target FER of $10^{-4}$, the SNR gap to the RCU bound is reduced to $0.1$ dB for the CRC-TBCC with $m=10$ and $v = 6$. Fig. \ref{Fig:TB_EL_v5_N128} shows the trade-off between the expected list rank $\E[L]$ and the FER. We see that the expected list rank $\E[L]<7$ for achieving the target FER of $10^{-4}$ for $v = 6$ and $m\le 10$, implying a low average decoding complexity of SLVD.

\section {Conclusion}
\label{sec: conclusion}
This paper shows that a rate-$(\omega-1)/\omega$ CC concatenated with a DSO CRC polynomial yields a good high-rate CRC-aided CC that approaches the RCU bound for the BI-AWGN channel. In particular, the best CRC-TBCCs approaches the RCU bound within $0.1$ dB for FER $10^{-4}$ at blocklength $N=128$ bits. Adding one bit to the CRC can improve the FER far more than adding an additional memory element to the CC does.

\section*{Acknowledgment}
The authors thank Dariush Divsalar for helpful discussions on construction of the dual trellis. We thank Ethan Liang and Linfang Wang for guidance and mentorship.  We also thank Mai Vu and Mohammad Karimzadeh for helpful collaboration.

% \section*{References}
\bibliographystyle{IEEEtran}
\bibliography{IEEEabrv,references}
% \printbibliography
% \end{thebibliography}

\end{document}